\documentclass[a4paper,11pt]{article}
\pdfoutput=1
\usepackage{a4wide}
\usepackage[english]{babel}
\usepackage{subfig}
\usepackage[T1]{fontenc} 
\usepackage{lmodern}
\usepackage{graphicx}
\usepackage{tabularx}
\usepackage{ltablex}
\usepackage{epsfig}
\usepackage{amsmath}
\usepackage{amssymb}
\usepackage{bm}
\usepackage{dsfont}
\usepackage{float}
\usepackage{placeins}
\usepackage{xcolor}
\usepackage{shuffle}
\usepackage{mathtools}
\usepackage{mathrsfs}
\usepackage{float}
\usepackage{shuffle}
\usepackage{cite}
\usepackage{hyperref}
\hypersetup{
    bookmarksnumbered,      
    bookmarksopen=true,     
    bookmarksopenlevel=3,   
    colorlinks,             
    allcolors=black,        
    citecolor=[HTML]007f00, 
    urlcolor=[HTML]00007f,  
    linkcolor=[HTML]b70000, 
    pdftitle={Elliptic Multiple Polylogarithms with Arbitrary Arguments in GiNaC},
}

\allowdisplaybreaks[4]
\DeclareMathOperator{\SL}{SL}

\newcommand{\im}{\text{Im}}
\newcommand{\re}{\text{Re}}
\newcommand{\diff}{\text{d}}
\newcommand{\Gt}{\tilde{\Gamma}}
\newcommand{\ssGt}{\tilde{\Gamma}\left( \begin{smallmatrix} n_1 & \dots & n_k \\ z_1 & \dots & z_k \end{smallmatrix}; z, \tau \right)}
\newcommand{\sm}[1]{\begin{smallmatrix} #1\end{smallmatrix}}

\newcommand{\cD}{\begin{cal}D\end{cal}}
\newcommand{\cF}{\begin{cal}F\end{cal}}
\newcommand{\cO}{\begin{cal}O\end{cal}}
\newcommand{\Z}{\mathds{Z}}
\newcommand{\N}{\mathds{N}}

\newcommand{\C}{\mathds{C}}
\newcommand{\HH}{\mathds{H}}

\newcommand{\ginac}{\texttt{GiNaC}}

\begin{document}

\thispagestyle{empty}

\begin{flushright}
BONN-TH/2026-01 \\ 
ZU-TH 01/26 \\
MITP-26-003
\end{flushright}

\vspace{1.5cm}

\begin{center}
  {\Large\bf Elliptic Multiple Polylogarithms with Arbitrary Arguments in \textsc{GiNaC}\\
  }
  \vspace{1cm}
  {\large Claude Duhr${}^{a}$, Florian Lorkowski${}^{b}$, Robin Marzucca${}^{b}$, Sofia Mauch${}^{c}$ and Stefan Weinzierl${}^{d}$} \\
  \vspace{1cm}
      {\small \em ${}^{a}$ Bethe Center for Theoretical Physics, Universität Bonn, D-53115 Bonn, Germany} \\
  \vspace{2mm}
      {\small \em ${}^{b}$ Physik-Institut, Universität Zürich, Winterthurerstrasse 190, 8057 Zürich, Switzerland} \\
  \vspace{2mm}
      {\small \em ${}^{c}$ ETH Zurich Wolfgang-Pauli-Str. 27, 8093 Zurich, Switzerland} \\
  \vspace{2mm}
      {\small \em ${}^{d}$ PRISMA Cluster of Excellence, Universit{\"a}t Mainz, D-55099 Mainz, Germany}\\ 
\end{center}

\vspace{2cm}

\begin{abstract}\noindent
  {
We present an algorithm for the numerical evaluation of elliptic multiple polylogarithms for arbitrary arguments and to arbitrary precision. The cornerstone of our approach is a procedure to obtain a convergent $q$-series representation of  elliptic multiple polylogarithms. Its coefficients are expressed in terms of ordinary multiple polylogarithms, which can be evaluated efficiently using existing libraries. In a series of preparation steps the elliptic polylogarithms are mapped into a region where the $q$-series converges rapidly. We also present an implementation of our algorithm into the \texttt{GiNaC} framework. This release constitutes the first public package capable of evaluating elliptic multiple polylogarithms to high precision and for arbitrary values of the arguments.
   }
\end{abstract}

\vspace*{\fill}


\newpage 

{\bf\large PROGRAM SUMMARY}
\vspace{4mm}
\begin{sloppypar}
\noindent 
   {\em Preliminary remark\/}: The algorithms are integrated into {\tt GiNaC}, this program summary gives first general information on {\tt GiNaC}, and then specific information on the sub-part related to the numerical evaluation of elliptic multiple polylogarithms with arbitrary arguments. \\[4mm]
{\bf General}: \\[2mm]
   {\em Title of program\/}: {\tt GiNaC} \\[2mm]
   {\em Version\/}: 1.8.10 \\[2mm]
   {\em Original authors\/}: C. Bauer, A. Frink, R. Kreckel \\[2mm]
   {\em Catalogue number\/}: \\[2mm]
   {\em Program obtained from\/}: {\tt https://www.ginac.de/} \\[2mm]
   {\em License\/}: GNU Public License \\[2mm]
   {\em Computers\/}: all \\[2mm]
   {\em Operating system\/}: all \\[2mm]
   {\em Program language\/}: {\tt C++     } \\[2mm]
   {\em Other programs called\/}: 
         CLN library, available from {\tt https://www.ginac.de/CLN}. \\[2mm]
   {\em External files needed\/}: none \\[2mm]
{\bf Specific}: \\[2mm]
   {\em Keywords\/}: Elliptic multiple polylogarithms, Feynman integrals.\\[2mm]
   {\em Nature of the physical problem\/}: 
         Numerical evaluation of elliptic multiple polylogarithms with arbitrary arguments. \\[2mm]
   {\em Method of solution\/}: 
         $q$-expansion and reduction to (ordinary) multiple polylogarithms.\\[2mm] 
   {\em Memory required to execute\/}: 
         Depending on the complexity of the problem. \\[2mm]
   {\em Restrictions on complexity of the problem\/}: 
         Only limited by the available memory and CPU time. \\[2mm]
   {\em Typical running time\/}:
         Depending on the complexity of the problem. \\[2mm]
   {\em E-mail\/}: {\tt cduhr@uni-bonn.de}, {\tt florian.lorkowski@physik.uzh.ch}, \\[2mm]
   \phantom{{\em E-mail\/}:} {\tt robin.marzucca@physik.uzh.ch}, {\tt smauch@ethz.ch}, {\tt weinzierl@uni-mainz.de}
\end{sloppypar}

\newpage

\section{Introduction}

Over the last decade, the LHC has collected large amounts of data from particle collisions, which has already led, or will lead, to experimental measurement accuracies of scattering processes at the percent-level~\cite{CMS:2024iaa,ATLAS:2024irg,CMS:2024myi,ATLAS:2025ifq}. To ensure that this data can be used optimally for phenomenological studies, we need to achieve similar precision in theoretical calculations. Accordingly, we should aim to calculate theoretical predictions for QCD processes at next-to-next-to-leading order (NNLO) or higher across the board. One focus of the high-energy particle physics community is the study of the Higgs boson and its interactions with other particles, and future collider projects~\cite{FCC:2018byv,FCC:2018evy,CEPCStudyGroup:2018ghi} are envisioned to operate as `Higgs factories'. Due to the interplay between the Higgs boson and the masses of other Standard Model particles, a thorough analysis of the Higgs boson requires us to track the functional dependence of scattering amplitudes on the masses of the heaviest particles of the theory.

The computation of higher-order corrections in quantum field theories is tightly connected to the evalutation of multiloop Feynman integrals. It is well known that large classes of Feynman integrals can be evaluated in terms of multiple polylogarithms~\cite{Remiddi:1999ew,Gehrmann:2000zt,Goncharov2007,goncharov2011multiple}, a class of of special functions defined as iterated integrals naturally attached to the punctured Riemann sphere. This class of special functions is extremely well understood: we do not only have a very good control over the analytic properties of these functions (see, e.g., refs.~\cite{Maitre:2005uu,Maitre:2007kp,Goncharov:2010jf,Brown:2009qja,Duhr:2011zq,Duhr:2012fh,Duhr:2019tlz}), but we also have various public implementations of these functions into public packages that allow one to obtain numerical results in a fast and reliable way~\cite{Gehrmann:2001jv,Gehrmann:2001pz,Vollinga:2004sn,Buehler:2011ev,Frellesvig:2016ske,Naterop:2019xaf}.

When calculating Feynman integrals with internal masses beyond one-loop order, one can quickly encounter geometries beyond the Riemann sphere--see ref.~\cite{Bourjaily:2022bwx} for a recent review or ref.~\cite{Bargiela:2025vwl} for an analysis of the function space of two-loop Feynman integrals--such as elliptic curves~\cite{Sabry:1962rge,Broadhurst:1993mw,Laporta:2004rb,Caron-Huot:2012awx,Adams:2013nia,Bloch:2013tra,Adams:2014vja,Remiddi:2016gno,Adams:2016xah,Broedel:2017siw,Kristensson:2021ani,Giroux:2022wav,Morales:2022csr,McLeod:2023qdf,Stawinski:2023qtw,Giroux:2024yxu,Spiering:2024sea}, hyperelliptic curves~\cite{Huang:2013kh,Marzucca:2023gto,Duhr:2024uid} or Calabi-Yau geometries~\cite{Brown:2010bw,Bloch:2014qca,Bloch:2016izu,Primo:2017ipr,Bourjaily:2018ycu,Bourjaily:2018yfy,Bonisch:2021yfw,Broedel:2021zij,Duhr:2022pch,Lairez:2022zkj,Pogel:2022yat,Pogel:2022ken,Pogel:2022vat,Duhr:2022dxb,Cao:2023tpx,Doran:2023yzu,McLeod:2023doa,Duhr:2023eld,Frellesvig:2023bbf,Klemm:2024wtd,Driesse:2024feo,Duhr:2024hjf,Frellesvig:2024zph,Frellesvig:2024rea,Duhr:2025ppd,Duhr:2025lbz,Maggio:2025jel,Brammer:2025rqo,e-collaboration:2025frv,Duhr:2025kkq,Pogel:2025bca,Duhr:2025xyy}. Recently, also first instances of Fano~\cite{Schimmrigk:2024xid,delaCruz:2025szs} and Del Pezzo geometries~\cite{Bargiela:2025vwl} have been uncovered. The occurrence of such geometries has long posed an obstruction to the analytic calculation of Feynman integrals which we have recently learned to overcome~\cite{Pogel:2022yat,Pogel:2022ken,Gorges:2023zgv,Pogel:2024sdi,Frellesvig:2024rea,Duhr:2024uid,Duhr:2025xyy,e-collaboration:2025frv,Bree:2025tug,Duhr:2025lbz,Chaubey:2025adn,Yang:2025ofz}. In particular, in the case of elliptic curves, a relevant class of special functions has by now been identified to consist of elliptic multiple polylogarithms~\cite{Brown:2011wfj,Broedel:2014vla,Broedel:2017kkb,EnriquezZerbini} and iterated integrals of (meromorphic) modular forms~\cite{ManinModular,2014arXiv1407.5167B,Adams:2017ejb,matthes_AMS,Broedel:2021zij}. While this allows us to calculate Feynman integrals involving such geometries analytically, we are still missing the means to evaluate these analytic results numerically in an efficient manner, and there is currently only one publicly available package to evaluate such classes of functions~\cite{Walden:2020odh}. 
This package is limited to the region of convergence of the series expansion of the integrand.
The numerical evaluation is, however, essential in order to make phenomenological predictions. In this paper we will partly address this shortcoming by introducing an algorithm to evaluate a certain class of elliptic multiple polylogarithms (i.e. the $\tilde{\Gamma}$-functions) numerically for arbitrary arguments and to arbitrary precision. This will be done with the help of $q$-expansions.

The paper is structured as follows. In section~\ref{sec:maths}, we will give a brief overview of the mathematical background necessary for this paper. In particular, we will introduce elliptic curves and the class of elliptic polylogarithms which can be evaluated with the algorithm presented here. In section~\ref{sec:algorithm}, we will explain our algorithm and how to prepare elliptic polylogarithms for a well-defined $q$-expansion that can be evaluated efficiently. In section~\ref{sec:usage}, we will show how to use the package in practice. In section~\ref{sec:applications}, we will discuss applications of the package and highlight interesting aspects regarding the choices users can make. Furthermore, we will show benchmark comparisons with other methods to evaluate elliptic polylogarithms numerically. Finally, in section~\ref{sec:conclusions}, we will conclude the paper. We include several appendices where we discuss the convergence of series and where we collect a summary of the commands needed to run our code.

\section{Mathematical Preliminaries}\label{sec:maths}

\subsection{Elliptic Curves and Tori}

An elliptic curve can be defined as the algebraic variety obtained as the zero set of a polynomial equation of the form
\begin{align}\label{eq:elliptic_pol}
y^2 = P_n(x)\,,  ~~ n \in \lbrace 3,4 \rbrace \,,
\end{align}
where $P_n$ is a polynomial in $x$ of degree $n$ with $n$ distinct roots. In the following we assume $n=4$ for concreteness, and we denote the four distinct roots of $P_4$ by $a_i$, $1\le i\le 4$. A distinctive feature of elliptic curves is the existence of a unique\footnote{It is unique up to normalisation.} holomorphic differential. Integrating this holomorphic differential between two zeroes of the polynomial $P_4$ gives the \emph{elliptic periods} $\psi_i$,
\begin{equation}\begin{split}
    \psi_1 &= 2 \int_{a_2}^{a_3} \frac{\diff x}{\sqrt{(x-a_1)(x-a_2)(x-a_3)(x-a_4)}} \,, \\
    \psi_2 &= 2 \int_{a_3}^{a_4} \frac{\diff x}{\sqrt{(x-a_1)(x-a_2)(x-a_3)(x-a_4)}} \,,
\end{split}\end{equation}
where we have parametrised the defining polynomial explicitly in terms of its roots $a_i$.

While elliptic curves typically manifest themselves in the context of Feynman integrals through a polynomial equation as in eq.~\eqref{eq:elliptic_pol}, for mathematical considerations another formulation is often more appropriate. In particular, every (complex) elliptic curve is isomorphic to the torus defined as a the quotient of the complex plane by the lattice spanned by its two periods.
Throughout this paper, we will only consider elliptic curves as complex tori.

\begin{figure}
\centering
\includegraphics[width=0.9\textwidth]{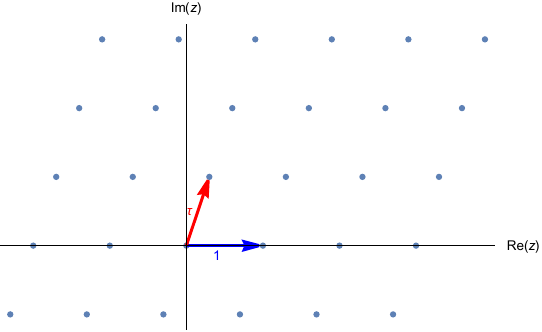}
\caption{The lattice $\Lambda_\tau$ spanned by the periods $\tilde{\psi}_1 = 1$ and $\tilde{\psi}_2 = \tau$. Each facet of the lattice corresponds to a copy of the torus. The whole complex $\C$ is the universal cover of the torus $\C/\Lambda_{\tau}$.}
\label{fig:lattice}
\end{figure}

When considering the torus as a topological object, we can rescale it by a non-zero factor without changing its geometry. We will use this fact to normalise the torus such that its periods are given by
\begin{align}
    \tilde{\psi}_1 = 1 ~~ \textrm{and} ~~ \tilde{\psi}_2 \equiv \tau = \frac{\psi_2}{\psi_1} \,,
\end{align}
where $\tau \in \HH^+$ lies in the upper-half-plane $\HH^+ = \lbrace a+ib \in \C \, | \, b > 0 \rbrace$ and is often referred to as the \emph{modular parameter}. We can thus represent the torus as the parallelogram in the complex plane spanned by the two periods $1$ and $\tau$ after identifying opposite sides. We may then consider the lattice $\Lambda_\tau = \Z+\tau\Z$ (see figure~\ref{fig:lattice}), and the torus can be described as the quotient $\C/\Lambda_{\tau}$.
The map that assigns to a point $(x,y)$ on the (affine) variety defined by eq.~\eqref{eq:elliptic_pol} the point $z \in \C$ is\footnote{Technically, the whole complex plane is the universal cover of the torus $\C/\Lambda_{\tau}$.}
\begin{align}
    z = \frac{1}{\psi_1}\int_{a_1}^x\frac{\diff x}{\sqrt{(x-a_1)(x-a_2)(x-a_3)(x-a_4)}} \,.
\end{align}

The periods spanning the lattice $\Lambda_{\tau}$ are not unique. It is easy to check, for example, that the periods $(1,\tau)$ and $(1,\tau +1)$ lead to the same lattice, and they hence define the same torus. The transformations that preserve this lattice are parametrised by the special linear group $\SL(2,\Z)$, which acts via Möbius transformations on $\tau$,
\begin{align}\label{eq:moebius}
    \tau \mapsto \tau' = \frac{a \tau + b}{c \tau + d} \, ~~ \text{for} ~~ \left(\begin{smallmatrix} a & b \\ c & d \end{smallmatrix}\right) \in \SL(2,\Z) \,.
\end{align}
For every $\tau\in\HH^+$ there is a M\"obius transformation in $\SL(2,\Z)$ that maps $\tau$ to a point $\tau'$ in the \emph{fundamental domain} for $\SL(2,\Z)$,
\begin{align}\label{eq:fundamental_domain}
    \mathcal{F} = \{\tau \in \HH^+: (|\tau|>1 \land-\tfrac{1}{2}\le \textrm{Re}(\tau)<\tfrac{1}{2})\vee(|\tau|=1\land -\tfrac{1}{2}\le \textrm{Re}(\tau)\le 0)\}\,.
\end{align}

\subsection{Multiple Polylogarithms and their Elliptic Analogues}\label{sec:empl}

Multiple polylogarithms (MPLs)~\cite{Goncharov2007,goncharov2011multiple} are a class of special functions that generalise of the classical logarithm and polylogarithm functions. They may be defined via the iterated integral,
\begin{equation}
    G(a_1,\ldots,a_n;x) = \int_0^x\frac{\diff t}{t-a_1}G(a_2,\ldots,a_n;t) \,, \label{eq:MPL_def}
\end{equation}
where the recursion starts with $G(;x)=1$, and the $a_i$ and $x$ are complex numbers. Whenever $a_n=0$, the integral in eq.~\eqref{eq:MPL_def} is divergent and needs to be regularised. We use the definition
\begin{equation}
    G(\underbrace{0,\ldots,0}_{n\textrm{ times}};x) = \frac{1}{n!}\log^nx\,.
\end{equation}
This regularisation corresponds to replacing the lower integration boundary by a so-called \emph{tangential base-point}, see ref.~\cite{2014arXiv1407.5167B}.

The main topic of this paper are analogues of MPLs associated with elliptic curves.
There are various definitions of elliptic multiple polylogarithms (eMPLs) in the literature~\cite{Brown:2011wfj,Broedel:2014vla,Adams:2014vja,Broedel:2017kkb}.
Here we consider the class of eMPLs~\cite{Brown:2011wfj,Broedel:2014vla} defined as iterated integrals over (purely meromorphic) integration kernels related to the \emph{Eisenstein-Kronecker series}~\cite{Brown:2011wfj,Kronecker:1881}, defined by
\begin{align}
    F(z,\alpha,\tau) = \frac{1}{\alpha} \sum_{n=0}^\infty g^{(n)}(z,\tau) \alpha^n =\frac{\theta_1'(0,\tau) \theta_1(z+\alpha,\tau)}{\theta_1(z,\tau) \theta_1(\alpha,\tau)} \,,
\end{align}
where $\theta_1$ is the odd Jacobi theta function and $\theta_1'(z,\tau) = \partial_z \theta_1(z,\tau)$.

The coefficients $g^{(n)}(z,\tau)$ of the series expansion of the Eisenstein-Kronecker series define the integration kernels of our class of eMPLs, defined recursively as
\begin{align}
    \ssGt &= \int_0^z \diff t \, g^{(n_1)}(t-z_1,\tau) \, \Gt \left(\sm{n_2 & \dots & n_k \\ z_2 & \dots & z_k} ; t, \tau \right)\,, \label{eq:def_Gt1}
\end{align}
where $\tau\in\HH^+$, $z$ and $z_i$ are arbitrary complex numbers. The recursion starts with $\Gt(;z,\tau)=1$, and we define the integration contour to be the straight line from $0$ to $z$, possibly deformed around poles.
We will refer to the arguments $z_i$ as the set of \emph{singular loci} and to the $n_i$ as the \emph{indices} of the eMPL. Just like ordinary MPLs, some instances require regularisation. We will discuss the regularisation below, after reviewing some of the properties of eMPLs and of the functions $g^{(n)}(z,\tau)$.

The class of functions defined by eq.~\eqref{eq:def_Gt1} frequently appears in the computation of Feynman integrals associated to elliptic curves. Hence, a good understanding of their analytic and numerical properties is paramount if we want to use eMPLs to evaluate multi-loop Feynman integrals relevant for phenomenology. A first step was taken in ref.~\cite{Walden:2020odh}, where an implementation of eMPLs into {\ginac} was introduced. This implementation, however, poses restrictions on the possible values of the $z_i$ and $z$. The main goal of this paper is to present an algorithm for their evaluation with arbitrary complex arguments, and to present the implementation of this algorithm into {\ginac}.

\paragraph{Properties of the integration kernels $\bm{g^{(n)}(z,\tau)}$.}
We briefly review the main properties of the integration kernels $g^{(n)}(z,\tau)$, because they form the basis to understand the properties of our eMPLs. The properties of the $g^{(n)}(z,\tau)$ derive from the properties of the Eisenstein-Kronecker series (cf.,~e.g.,~refs.~\cite{Kronecker:1881,Brown:2011wfj,Broedel:2014vla}). For example, the Eisenstein-Kronecker series is quasi-periodic,
\begin{align}\label{eq:F_periodicity}
    F(z+1,\alpha,\tau) &= F(z,\alpha,\tau)\,, & F(z+\tau,\alpha,\tau) &= e^{-2 \pi i \alpha}F(z,\alpha,\tau)\,,
\end{align}
which translates to the coefficients $g^{(n)}(z,\tau)$ as
\begin{align}
    g^{(n)}(z+1,\tau) &= g^{(n)}(z,\tau) \,,                                    \label{eq:g_periodicity1} \\
    g^{(n)}(z+ \tau,\tau) &= \sum_{k=0}^n (- 2 \pi i)^k g^{(n-k)}(z,\tau) \,.   \label{eq:g_periodicity2}
\end{align}
Since the $g^{(n)}$ are only quasi-periodic, they are strictly speaking not well-defined functions on the torus, but rather on its universal cover, which is the complex plane $\C$.
Similarly, the class of eMPLs defined through eq.~\eqref{eq:def_Gt1} is only defined on the universal cover. A periodic version of eMPLs defined on the torus is often found in the mathematics and string theory literature~\cite{Brown:2011wfj,Broedel:2014vla}.

The functions $g^{(n)}(z,\tau)$ also have definite parity,
\begin{align}
    g^{(n)}(-z,\tau) = (-1)^n g^{(n)}(z,\tau) \,.
\end{align}
Moreover, they are meromorphic functions of $z$. In particular,
$g^{(1)}(z,\tau)$ has a simple pole at every point of the lattice $\Lambda_{\tau}$. The functions $g^{(n)}(z,\tau)$ with $n \geq 2$ are finite on the real axis and, by virtue of the quasi-periodicity in eq.~\eqref{eq:g_periodicity1}, exhibit simple poles at other lattice points through the kernels $g^{(1)}$. Note that this implies that the eMPLs~$\Gt$ defined in eq.~\eqref{eq:def_Gt1} exhibit at most logarithmic singularities in $z$.
Finally, the functions $g^{(n)}(z,\tau)$ also have a well-defined behaviour under modular $\SL(2,\Z)$ transformations as in eq.~\eqref{eq:moebius}:
\begin{align}\label{eq:g_modular}
    g^{(n)}\!\left(\frac{z}{c\tau+d},\frac{a\tau+b}{c\tau+d}\right) &\,= (c\tau+d)^n\,\sum_{k=0}^n\frac{1}{k!}\,\left(\frac{2\pi i\, c\,z}{c\tau+d}\right)^k\,g^{(n-k)}(z,\tau) \,.
\end{align}

Since our goal is to present an algorithm to evaluate eMPLs numerically, a first question is how to evaluate the integration kernels $g^{(n)}(z,\tau)$. One efficient way to do so is to expand them into a so-called \emph{$q$-series}, i.e., an expansion in the parameter:
\begin{align}
    q_\tau \equiv \exp (2 \pi i \tau) \,. \label{eq:qtau}
\end{align}
Explicitly, we have
\begin{align}
    g^{(1)}(z,\tau) &= \pi  \cot(\pi z) + 4\pi \sum_{m=1}^{\infty}  \sin(2\pi m z) \sum_{n=1}^{\infty} q_\tau^{mn} \,, \nonumber \\
    g^{(k)}(z,\tau) \Big|_{k=2,4,\ldots} &= - 2 \zeta_k -2 \frac{ (2\pi i)^k }{(k-1)!} \sum_{m=1}^{\infty}  \cos(2\pi m z) \sum_{n=1}^{\infty}n^{k-1} q_\tau^{mn} \,, \label{eq:gQExpansion} \\
    g^{(k)}(z,\tau) \Big|_{k=3,5,\ldots} &= - 2i \frac{ (2\pi i)^k }{(k-1)!} \sum_{m=1}^{\infty}  \sin(2\pi m z) \sum_{n=1}^{\infty}n^{k-1} q_\tau^{mn} \,, \nonumber
\end{align}
and $\zeta_k = \sum_{n\ge 1}n^{-k}$ denotes the Riemann zeta function.
Note that the absolute value of $q_\tau$ only depends on the imaginary part of the modular parameter $\tau$. As we have seen, we can always describe a torus with a parameter $\tau$ in the fundamental domain $\mathcal{F}$ in eq.~\eqref{eq:fundamental_domain}. This implies that we can always ensure via a modular transformation that $\im(\tau) > \sqrt{3}/2$, and thus $|q_\tau| \lesssim 0.004$. The convergence of the double sums in eq.~\eqref{eq:gQExpansion} does however not only depend on the value of $\tau$, but also on the value of $z$, and convergence is guaranteed as long as $|\im(z)| < \im(\tau)$. A proof of this statement can be found in appendix~\ref{app:convergence}.

\paragraph{Properties of eMPLs.} We now review some of the properties of the eMPLs defined by eq.~\eqref{eq:def_Gt1}. First of all, since eMPLs are defined as iterated integrals, they enjoy all the general properties of iterated integrals~\cite{Chen:1977oja}. In particular, they form a shuffle algebra, and we have
\begin{align}
    \Gt\left(\sm{n_1&\dots&n_k\\z_1&\dots&z_k};z,\tau\right)\Gt\left(\sm{n_{k+1}&\dots&n_{k+l}\\z_{k+1}&\dots&z_{k+l}};z,\tau\right) &= \sum_{\mathclap{\sigma \in \Sigma(k,l)}} \Gt\big(\sm{n_{\sigma(1)}&\dots&n_{\sigma(k+l)} \\ z_{\sigma(1)}&\dots&z_{\sigma(k+l)}};z,\tau\big)\,,
\end{align}
where $\Sigma(k,l)$ is the set of all shuffles of two sets of $k$ and $l$ elements. They also satisfy the path decomposition formula, which states that for two paths $\gamma_1, \gamma_2$ such that $\gamma_1$ ends in the beginning of $\gamma_2$, we have
\begin{align}\label{eq:path_composition}
    \int_{\gamma_1\gamma_2} g_1 \dots g_k = \sum_{i=0}^k \int_{\gamma_2} g_1 \dots g_i \int_{\gamma_1} g_{i+1} \dots g_k \,,
\end{align}
where $g_i$ denotes the $i^{\textrm{th}}$ integration kernel in the definition~\eqref{eq:def_Gt1}, and where, for $\gamma = [0,z]$,
\begin{align}
    \int_{\gamma} g_1 \dots  g_k = \ssGt
\end{align}
denotes the iterated integration of the kernels $g_i$ along the path $\gamma$.

Second, there are properties that directly follow from the corresponding properties of the integration kernels $g^{(n)}(z,\tau)$. For example, one can work out the behaviour of eMPLs under translations by the lattice $\Lambda_{\tau}$ and $\SL(2,\Z)$ transformations. Moreover, since the integration kernels $g^{(n)}(z,\tau)$ have simple poles in $z$, we see that eMPLs may exhibit logarithmic divergences.
From this we can infer that the integral representation in eq.~\eqref{eq:def_Gt1} becomes ill-defined whenever $\left(\sm{n_k\\z_k}\right)= \left(\sm{1\\0}\right)$, and these cases require regularisation. In the case where we only have the kernel $ \left(\sm{1\\0}\right)$, we define
\begin{align}
    \Gt \left( \sm{1 \\ 0} ; z , \tau \right) &= \log ( 1 - e^{2 \pi i z}) - 2 \pi i z + \int_0^z \diff t \left( g^{(1)}(t,\tau) - \frac{2 \pi i}{e^{2 \pi i t} - 1} \right)\,. \label{eq:Gt10}
\end{align}
Note that the regularisation in this form only holds if the exponential $\exp (2 \pi i z t)$, with $t \in [0,1]$, does not trace out a contour which crosses the line extending from $-1$ to $-\infty$. This is precisely the case if $\im(z) > 0 $ or if $ \re (z) \in (-0.5,0.5)$. In such cases, the logarithm appearing in eq.~\eqref{eq:Gt10} is not on its principle branch, and a multiple of $2 \pi i$ needs to be added. For integrals consisting of multiple $\left(\sm{1\\0}\right)$ kernels, we define
\begin{align}
    \Gt\big( \underbrace{\sm{1 & \dots & 1 \\ 0 & \dots & 0}}_{n \text{ times}}; z,  \tau \big) &= \frac{\Gt \left( \sm{1 \\ 0} ; z , \tau \right)^n}{n!} \,, \label{eq:Gt10all}
\end{align}
while for general eMPLs ending in $\left(\sm{1\\0}\right)$, we use the relation~\cite{Panzer:2015ida}
\begin{align}\label{eq:regularisation_Panzer}
    \Gt\big( \sm{n_1&\dots&n_k\\z_1&\dots&z_k} \underbrace{\sm{1&\dots&1\\0&\dots&0}}_{n \text{ times}}; z, \tau \big) &=
        \sum_{i=0}^n (-1)^i \Gt\big( ( \sm{n_1&\dots&n_{k-1}\\z_1&\dots&z_{k-1}} \shuffle \underbrace{\sm{1&\dots&1\\0&\dots&0}}_{i \text{ times}} ) \sm{n_k\\z_k}; z,\tau \big) \frac{\Gt\left(\sm{1\\0};z,\tau\right)^{n-i}}{(n-i)!}\,,
\end{align}
with $\left(\sm{n_k\\z_k}\right) \neq \left(\sm{1\\0}\right)$. This regularisation procedure corresponds to the tangential base point regularisation described in ref.~\cite{Broedel:2014vla}.

\section{Numerical Evaluation of eMPLs via \texorpdfstring{$\bm{q}$}{q}-series}\label{sec:algorithm}

In this section we introduce a method for the efficient numerical evaluation of elliptic multiple polylogarithms. In a nutshell, our method relies on the following result, which is also one of the main theoretical results of our paper:
\begin{quote}\emph{
Every eMPLs admits a convergent $q$-series representation of the form
\begin{equation}
    \ssGt = \sum_{m=0}^\infty q_{\tau'}^m\,C_m(\{z_i\};z,\tau') \,, \label{eq:Gt_q_series}
\end{equation}
where the $C_m(\{z_i\},z;\tau')$ are combinations of ordinary MPLs and rational functions in suitable variables.
The variable $\tau'$ is related to $\tau$ by a modular transformation.
}
\end{quote}
As a consequence of this statement, we obtain an efficient method to evaluate eMPLs numerically by truncating the infinite sum at some order and evaluating the coefficients $C_m(\{z_i\},z;\tau')$ using well-established libraries for the evaluation of ordinary MPLs, cf.,~e.g., refs.~\cite{Gehrmann:2001jv,Gehrmann:2001pz,Vollinga:2004sn,Buehler:2011ev,Frellesvig:2016ske,Naterop:2019xaf}.
The proof of this claim will be provided in the remainder of this section as an algorithm to evaluate the coefficients $C_m(\{z_i\},z;\tau')$ analytically in terms of ordinary MPLs. In essence, the idea consists in inserting the $q$-series representation of the integration kernels from eq.~\eqref{eq:gQExpansion} into the integral representation of eMPLs in eq.~\eqref{eq:def_Gt1} and then exchanging the summation and the integration.  However, special care is required, because we need to make sure to obtain a convergent series representation after the sums and the integrals have been exchanged. Moreover, MPLs and eMPLs are multivalued functions, and we need to correctly capture the branch cut structure of the transcendental functions.
Our algorithm involves multiple preparation steps, in which the eMPLs are rewritten in a form more suitable for computing the $q$-series, and thus numerical evaluation. Eventually, a $q$-expansion is performed that is capable of evaluating eMPLs to arbitrary precision. A summary of the different steps is:

\begin{enumerate}
    \item Shift singular loci to a fundamental domain. \label{algo:step1}
    \item Apply regularisation of singular kernels. \label{algo:step2}
    \item (Optional) Map the modular parameter to its fundamental domain, repeat steps \ref{algo:step1} and \ref{algo:step2}. \label{algo:step3}
    \item Decompose and shift integration contour, repeat step \ref{algo:step1}. \label{algo:step4}
    \item Perform $q$-expansion. \label{algo:step5}
\end{enumerate}
The steps~\ref{algo:step3} and~\ref{algo:step4} will need to be executed at most once. In the remainder of this section we describe these different steps in detail, and in the next section we present an implementation of our algorithm into {\ginac}.

\subsection{Shifting the Singular Loci}

We have seen that the integration kernels $g^{(n)}(z,\tau)$ can exhibit poles at the lattice points $z\in \Lambda_{\tau}$, and they are quasi-periodic with respect to shifts by the lattice $\Lambda_{\tau}$.
A first preparatory step of our algorithm consists in using the quasi-periodicity to shift all the singular loci to a specific connected region $\cD\subset \C$, which we call the the \emph{fundamental domain of singular loci}.
We may apply eqs.~\eqref{eq:g_periodicity1} and~\eqref{eq:g_periodicity2} under the integral sign in the definition of the eMPLs in eq.~\eqref{eq:def_Gt1}. This allows us to replace each eMPL by a linear combination of eMPLs whose singular loci all lie inside our chosen fundamental domain $\cD$.

At this point, there is some arbitrariness in how we choose the fundamental domain $\cD$, because for example we could still translate it by the lattice $\Lambda_{\tau}$.
For our purposes, we choose $\cD$ as the region bounded by $\pm \tfrac{1}{2}$ and $\pm\tfrac{1}{2} \im(\tau)$:
\begin{equation}
    \cD = \big\{z\in \C : -\tfrac{1}{2}\le \re(z) < \tfrac{1}{2} \land -\tfrac{1}{2}\im(\tau)\le \im(z) < \tfrac{1}{2}\im(\tau)\big\}\,.
\end{equation}
Our specific choice is motivated by the following reasons. First, the appearance of singular $\left(\sm{1\\0}\right)$ kernels in the iterated integrals has become manifest (because the only singular locus in $\cD$ that leads to a divergent integral is the point $0\in\cD$), making it clear which eMPLs require regularisation. Second, if the integration contour lies within $\cD$, the $q$-series in eq.~\eqref{eq:gQExpansion} converges on the whole integration contour. Third, the singular loci in the convergent integration domain are manifest, i.e., the singular loci are explicitly defined by a complex number in the convergent integration domain and not by an (equivalent) value shifted by a lattice point.\footnote{Note that the integration kernels in the iterated integral appear as $\diff t_i \,g^{(n_i)}(t_i-z_i,\tau)$, making the $z_i$ as singular loci obvious.}

Note that the convergence of the $q$-series in eq.~\eqref{eq:gQExpansion} depends only on the imaginary part of the arguments, and we would not need to limit the convergence domain in the real direction. The restriction to real parts between $\pm 0.5$ is due to the periodicity of the trigonometric functions appearing in the expansion. We will see that the arguments of the polylogarithms will be of the form $q_z = \exp (2 \pi i z)$. Numerically, the expressions $q_z$ and $q_{z+1}$ are indistinguishable. The formal difference between the two will, however, be relevant if they appear as the argument of a simple logarithm, $\log q_z$, as they would lie on different Riemann sheets. These logarithms appear as an artefact of the change of variables we will need to perform (cf.~eq.~\eqref{eq:cov}) to rationalise the trigonometric functions appearing in eq.~\eqref{eq:gQExpansion}. To avoid having to track the winding of the arguments around the logarithmic poles, we restrict ourselves to the domain $\cD$ bounded by $\pm 1/2$, as this choice corresponds to the branch of the logarithm that is conventional in most mathematical implementations, including {\ginac}. If our algorithm is used in conjunction with other numerical libraries to evaluate ordinary MPLs, care must be taken to match the branches of the logarithm implemented in those libraries.

\subsection{Regularisation of eMPLs}

After the first preparatory step, we may assume that all singular loci of all eMPLs lie in the fundamental domain $\cD$. As a consequence, the only eMPLs that require regularisation are those whose last letter is $\left(\sm{1\\0}\right)$. We may then apply eqs.~\eqref{eq:Gt10all} and~\eqref{eq:regularisation_Panzer}. In this way we manage to reduce the eMPLs that require regularisation entirely to eq.~\eqref{eq:Gt10}. Hence, at the end of this step, we may assume that all eMPLs admit a convergent representation in terms of iterated integrals and that there are no eMPLs left that require regularisation. Regularised eMPLs require special treatment in subsequent steps. By applying the regularisation procedure described in this step, we limit the cases that need to be treated separately to the eMPL in eq.~\eqref{eq:Gt10}.

\subsection{Mapping the Modular Parameter to the Fundamental Domain}

We eventually want to insert the $q$-series of the integration kernels in eq.~\eqref{eq:gQExpansion} into the representation of eMPLs as iterated integrals in eq.~\eqref{eq:def_Gt1}. If the singular loci lie in the fundamental domain $\cD$, then the $q$-series will converge as long as $\im(\tau) > 0$ (see appendix~\ref{app:convergence}). We may, however, perform a modular transformation to change the value of $\tau$ such as to improve the numerical convergence of the $q$-series.

This is the purpose of step 3. Note that, strictly speaking, this step is optional, and it is not mandatory to obtain a convergent $q$-series as in eq.~\eqref{eq:Gt_q_series}. It may, however, greatly improve the numerical convergence of the $q$-series. We know that we can always find a modular transformation as in eq.~\eqref{eq:moebius} such that $\tau'$ lies in the fundamental domain $\mathcal{F}$ for $\SL(2,\Z)$ in eq.~\eqref{eq:fundamental_domain}. We also know how the integration kernels transform under modular $\SL(2,\Z)$ transformations (cf.~eq.~\eqref{eq:g_modular}). If we apply the modular transformation property eq.~\eqref{eq:g_modular} to the integration kernels as they appear in the iterated integral, we have
\begin{align}\label{eq:g_modular_2}
    g^{(n)}\!\left(t - z,\tau\right) &\,= (\gamma\tau'+\delta)^{n}\,\sum_{k=0}^n\frac{1}{k!}\,\left(\frac{2\pi i\, \gamma\,(t' - z')}{\gamma\tau'+\delta}\right)^k\,g^{(n-k)}(t' - z',\tau') \,,
\end{align}
where $z' = (\gamma\tau'+\delta)z$ and $t' = (\gamma\tau'+\delta)t$ are the mapped singular locus and the mapped integration variable respectively and $\gamma$ and $\delta$ are the parameters of the inverse Möbius transform $\left(\begin{smallmatrix}\alpha&\beta\\\gamma&\delta\end{smallmatrix}\right) = \left(\begin{smallmatrix}a&b\\c&d\end{smallmatrix}\right)^{-1}$.
This relation enables us to rewrite each integration kernel as a linear combination of kernels whose modular parameter lies in the fundamental domain $\cF$ and whose prefactors are polynomials in the integration variable. To write these again as eMPLs, we iteratively integrate each term, starting with the innermost kernels. At each step, the integrand consists of an eMPL, an integration kernel $g^{(n)}$ and/or powers of the integration variable $t$. These integrals can be performed using the formulas (which are nothing but a recasting of the definition in eq.~\eqref{eq:def_Gt1}):
\begin{align}
    \int_0^z \diff t \, g^{(n_1)}(t-z_1,\tau)                                                          &= \Gt\left(\sm{  n_1          \\  z_1          };z,\tau\right) \,, \label{eq:int1} \\
    \int_0^z \diff t \,                       \Gt\left(\sm{n_2&\dots&n_k\\z_2&\dots&z_k};t,\tau\right) &= \Gt\left(\sm{0&n_2&\dots&n_k\\0&z_2&\dots&z_k};z,\tau\right) \,, \label{eq:int2} \\
    \int_0^z \diff t \, g^{(n_1)}(t-z_1,\tau) \Gt\left(\sm{n_2&\dots&n_k\\z_2&\dots&z_k};t,\tau\right) &= \Gt\left(\sm{  n_1&\dots&n_k\\  z_1&\dots&z_k};z,\tau\right) \,, \label{eq:int3}
\end{align}
as well as the integration-by-parts identity
\begin{align}
    \int_0^z \diff t \, t^n f(t) &= z^n \int_0^z \diff t \, f(t) - n \int_0^z \diff t \, t^{n-1} \int_0^t \diff u \, f(u) \,,
\end{align}
where $f(t)$ can be any of the integrands from eqs.~\eqref{eq:int1}, \eqref{eq:int2} and \eqref{eq:int3}.
After having performed this step, we may assume that all eMPLs have an argument $\tau$ that lies in the fundamental domain $\mathcal{F}$. In particular, this means that after this step, we may assume that $\im(\tau) \ge \sqrt{3}/2$ and thus $|q_\tau| \lesssim 0.004$, which ensures a rapid convergence of the $q$-expansion, as the convergence will depend on $|q_\tau|$.

Let us conclude the discussion of this step by making some comments.
First, when we apply eq.~\eqref{eq:g_modular_2}, we need to rescale the singular loci $z \mapsto z'$, possibly resulting in values outside of the fundamental domain $\cD$. This necessitates redoing the first preparation step of the algorithm, and to shift them back into the fundamental domain. These shifts can map the singular loci to zero, thereby creating new eMPLs that require regularisation. Thus, also the regularisation step needs to be redone.
Both of these transformations leave the modular parameter unchanged, such that step 3 does not need to be iterated. Note that it is not possible to first map the modular parameter to the fundamental domain before having executed steps 1 and 2, because the modular transformations must be performed on convergent integrals that do not require regularisation. Indeed,
a special treatment is necessary for eMPLs that require regularisation. We can check that for the regularisation procedure defined through eq.~\eqref{eq:Gt10}, the following relation holds.
\begin{align}
    \Gt\left(\sm{1\\0};z,\tau\right) &= \Gt\left(\sm{1\\0};z',\tau'\right) - \log(c\tau'+d) + \frac{i\pi c z'^2}{c\tau'+d} \,.
\end{align}

\subsection{Decomposing the Path of Integration}\label{subsec:cutIntegrationPath}

At this point, we have transformed a given eMPL into a combination of eMPLs such that all singular loci lie in the fundamental domain $\cD$. The upper integration limit $z$, however, is not constrained to lie in $\cD$ for now.
In order to ensure convergence of the $q$-series along the full integration path, we need to require that $|\im(z)| < \im(\tau)$. We can make sure that this condition is met by restricting not only the singular loci, but also the upper integration limit $z$ to the fundamental domain $\cD$.\footnote{Technically, the convergence only depends on the values of the imaginary parts. The restriction in the real direction ensures that we lie on the principle branch of the logarithm.} To achieve this, we can split the path into shorter segments using the path composition formula in eq.~\eqref{eq:path_composition} and map the lower boundary back to 0, so that all resulting eMPLs are defined as an iterated integral of a segment $(0,z_*)$, for some $z_*\in \cD$.

At this point we need to make a technical comment:
Restricting the singular loci and the upper integration limit $z$ to $\cD$ only formally guarantees convergence. As we have seen, the convergence of the $q$-expansions of the integration kernels $g^{(n)}(t-z_i,\tau)$, where $t$ is the integration variable and $z_i$ is the singular locus, requires the argument $t - z_i$ to satisfy $|\im(t-z_i)| < \im (\tau)$ on the whole integration domain. As the integration path is a straight line from $0$ to $z$, it suffices to ensure that the convergence criterion is fulfilled at the endpoints. Since $z_i \in \cD$, the convergence is guaranteed for $t = 0$, and we only need to ensure that $|\im (z-z_i)| < \im (\tau)$. If we allow $z \in \cD$, this is always the case, but we could in principle have $\im (z_i) = - \im(\tau) /2$ and $\im(z) \rightarrow \im(\tau)/2$, in which case the convergence would become infinitely slow. We therefore restrict the integration boundary $z$ to a smaller domain $\cD'\subseteq \cD$, which is given by
\begin{align}\label{eq:fundamentalDomain_d2}
    \cD' = \lbrace z \in \C \, : \, |\re(z)| \leq r_\text{max} \land |\im(z)| \leq i_\text{max} \im(\tau) \rbrace \,,
\end{align}
for some fixed $r_\text{max},i_\text{max} \in (0,1/2)$. We choose the default values $r_\text{max}=i_\text{max}=0.4$. Then the worst possible case converges as
\begin{align}
\sum_{k=0}^\infty \exp\left(- 2 \pi \, (0.5 - i_\text{max}) \, \im (\tau) \right)^k \,.
\end{align}
As expected, we see that the convergence worsens as $i_\text{max} \rightarrow 1/2$ and improves as $i_\text{max} \rightarrow 0$. We can also see that the convergence is (in a sense) proportional to $\im(\tau)$, and mapping $\tau$ to the fundamental domain generally improves convergence. We should also note that there is a trade-off when choosing smaller values $i_\text{max}$: On the one hand, the worst possible convergence improves, meaning that we will potentially need to integrate over fewer orders in the $q$-series. On the other hand, smaller values $i_\text{max}$ are realised by cutting the integration path into smaller pieces, generally yielding a larger number of prepared eMPLs which need to be expanded. Depending of the arguments of the eMPLs and the desired accuracy, the optimal value of $i_\text{max}$ may vary.

By default, we cut the integration contour into equidistant pieces. We find the smallest number of cuts, $n_\text{cuts}$, such that the cut and shifted contours are completely within $\cD'$ and the endpoints of the shifted contours do not coincide with poles in the integrand.
The division into equidistant segments ensures that the worst possible convergence of all resulting prepared eMPLs is equal. Furthermore, all prepared eMPLs will have the same argument $z/n_\text{cut}$, allowing for possible cancellations.

After having identified the appropriate decomposition of the original integration path $(0,z)$, we apply the path decomposition formula in eq.~\eqref{eq:path_composition} repeatedly to split the integration into as many paths as are required to ensure that all integrations are contained in the region $\cD'$. To be more precise, let $\gamma$ represent a path of $n$ segments, delimited by the set of points $c$. In particular, the $i^{\textrm{th}}$ segment of $\gamma$ stretches from $c_i$ to $c_{i+1}$. The endpoints are those of the original integration, i.e., $c_1=0$ and $c_{n+1}=z$. We can rewrite an eMPL that is integrated along this path as a combination of eMPLs with straight integration paths starting at the origin
\begin{align}
    \Gt\left(\sm{n_1&\dots&n_k\\z_1&\dots&z_k};z,\tau\right)\big\vert_\gamma &= \, \sum_{\mathclap{p \in \text{IP}(k,n)}} \quad \prod_{i=1}^{n} \, \Gt\left( \sm{
        n_{P_i}          &\dots&n_{P_{i+1}} \\
        z_{P_i}-c_{n-i+1}&\dots&z_{P_{i+1}}-c_{n-i+1}}
        ; c_{n-i+2}-c_{n-i+1}, \tau \right) \,.
\end{align}
Here, $p$ runs over the (ordered) integer partition of $k$ into $n$ elements, i.e., over the set of all tuples $(p_1,\ldots,p_n)$ with $p_i \in \N_0$ and $\sum_{i=1}^n p_i = k$ and we introduced the shortcut $P_i = 1 + \sum_{j=1}^{i-1} p_j$.

As this procedure shifts the singular loci, we again repeat the first step of the algorithm, to ensure that all singular loci fall in the fundamental domain. Since we ensured that none of the cuts of the path decomposition coincide with any of the singular loci, the regularisation step does not need to be repeated.

\subsection[The \texorpdfstring{$q$}{q}-Expansion of eMPLs]{The $\bm{q}$-Expansion of eMPLs}\label{sec:nqexpand}

After having performed the previous steps, we have obtained a combination of eMPLs with the following properties:
\begin{itemize}
\item The only eMPL that requires regularisation is $\Gt \left( \sm{1 \\ 0} ; z , \tau \right)$.
\item The modular parameter $\tau$ lies in the fundamental domain $\mathcal{F}$ for $\SL(2,\Z)$.
\item All singular loci lie inside the fundamental domain $\cD$.
\item All upper integration limits lie inside the domain $\cD' \subseteq \cD$.
\end{itemize}
These properties guarantee that we can expand the integration kernels into a convergent $q$-series. We now describe how we can perform this expansion and do all integrals order by order in the $q$-series in terms of ordinary MPLs and rational functions.

Let us denote by $t_j$ the integration variables in the iterated integration, and $z_j$ are the singular loci. It is convenient to introduce the variables
\begin{equation}
    w_j = \exp(2\pi i t_j) \textrm{~~and~~~} q_{z_j} = \exp(2\pi i z_j) \,. \label{eq:cov}
\end{equation}
The $q$-series for the integration kernels from eq.~\eqref{eq:gQExpansion} can then be cast in the form:
\begin{align}
    g^{(1)}(t_j-z_j,\tau) &= i\pi \frac{w_j+q_{z_j}}{w_j-q_{z_j}} + 2\pi i \sum_{m=1}^{\infty} \left( \frac{q_{z_j}}{w_j}-\frac{ w_j}{q_{z_j}} \right) \sum_{n=1}^{\infty} q_\tau^{mn} \,, \nonumber \\
    g^{(k)}(t_j-z_j,\tau)\Big|_{k=2,4,\ldots} &= -2\zeta_k - \frac{(2\pi i)^k}{(k-1)!} \sum_{m=1}^{\infty} \left[ \left(\frac{q_{z_j}}{w_j}\right)^m + \left(\frac{ w_j}{q_{z_j}}\right)^m \right] \sum_{n=1}^{\infty}n^{k-1} q_\tau^{mn} \,, \label{eq:expandedkernels}\\
    g^{(k)}(t_j-z_j,\tau)\Big|_{k=3,5,\ldots} &=             \frac{(2\pi i)^k}{(k-1)!} \sum_{m=1}^{\infty} \left[ \left(\frac{q_{z_j}}{w_j}\right)^m - \left(\frac{ w_j}{q_{z_j}}\right)^m \right] \sum_{n=1}^{\infty}n^{k-1} q_\tau^{mn} \,. \nonumber
\end{align}
We note at this point that, since $|q_\tau|<1$, it is convenient to perform the innermost sum explicitly and replace it by a finite sum, because it does not depend on the integration variables $w_j$,
\begin{equation}\label{eq:qSum}\begin{split}
    \sum_{n=1}^\infty q_{\tau}^n &= \frac{q_{\tau}}{1-q_{\tau}} \,, \\
    \sum_{n=1}^\infty n^k q_{\tau}^n &= \frac{1}{(1-q_{\tau})^{k+1}} \sum_{i=1}^k q_{\tau}^i \sum_{j=0}^{i-1} (-1)^j (i-j)^k \binom{k+1}{j} \,.
\end{split}\end{equation}

Since both the $q$-series in $q_{\tau}$ and the integrals over $w_j$ are absolutely convergent,\footnote{The absolute convergence of the integrals follows from the fact that they do not require regularisation. In cases where the integration contour runs over a singular locus, the contour is understood to be deformed around the pole.} we may exchange the summation and the integrations. We can then perform all the integrals in $w_j$ order by order in the $q$-series. From eq.~\eqref{eq:expandedkernels} we see that the coefficients of the series in $q_{\tau}$ are rational functions in the integration variables $w_j$. Hence, we may perform all integrations in terms of ordinary MPLs and rational functions. This finishes the proof of our claim that every eMPL admits a convergent $q$-series whose coefficients can be expressed in terms of ordinary MPLs and rational functions, cf.~eq.~\eqref{eq:Gt_q_series}. We obtain in this way an effective algorithm to evaluate eMPLs numerically, because the ordinary MPLs that appear in the coefficients can be evaluated using standard techniques~\cite{Gehrmann:2001jv,Gehrmann:2001pz,Vollinga:2004sn,Buehler:2011ev,Frellesvig:2016ske,Naterop:2019xaf}.

Due to the regularisation procedure for the formally divergent integral $ \Gt\left(\sm{1\\0};z,\tau\right)$, we need to take special care when calculating its $q$-expansion. In order to obtain the correct value in accordance with the tangential base-point prescription, we essentially follow the prescription given in eq.~\eqref{eq:Gt10}: We subtract the pole at the lower integration boundary from the $q$-expansion of the integration kernel and add the appropriate logarithms to the integrated expression.

\begin{figure}[t]
    \centering
    \includegraphics[page=1]{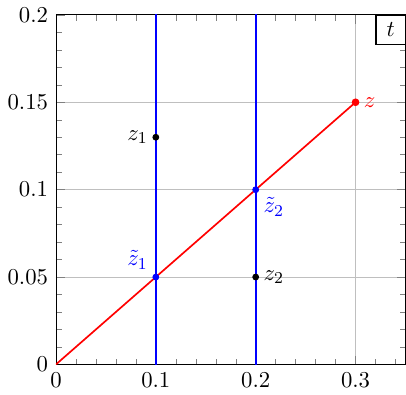}%
    \hfill%
    \includegraphics[page=2]{figures/coordinate_transformation.pdf}
    \caption{Integration contour in $t$-space (left) and $w$-space (right). The path shown in a solid red line corresponds to the integration in the original eMPL. The canonical path in $w$-space is shown as a dashed line. The image $w_1$ of the singular locus $z_1$ obstructs us from deforming the original contour to the dashed path. The orange segments indicate the integration path that we construct in $w$-space, which passes on the correct sides of all singular loci. The dashed orange line indicates the optimised integration path, which avoids adding a node at $\tilde{w}_2$.}
    \label{fig:path}
\end{figure}

At this point, we need to make a technical comment on how we can actually perform the integration in terms of ordinary MPLs in practise.
By convention, the path of integration in the definition of both ordinary and elliptic MPLs in eqs.~\eqref{eq:MPL_def} and~\eqref{eq:def_Gt1} is usually understood to be the straight line from $0$ to the argument $z$.\footnote{The only exception from this rule being the regularised versions of the divergent integrals $G(0;z)$ and $\Gt\left(\sm{1\\0};z,\tau\right)$.}
When changing variables from $t_i$ to $w_i$, the exponential function will in general map out a curved path in $w$-space (see figure~\ref{fig:path}).
This path can in general not be deformed into a straight line without changing the value of the integral, as this deformation might cross singular loci which can be poles in the integrand.

To avoid this issue, we express the path as a series of straight line segments. The construction of these segments is illustrated in figure~\ref{fig:path}. For each singular locus $z_i$, we construct a node $\tilde{z}_i$ which lies on the integration path and whose real part is identical to that of $z_i$. These new nodes are then mapped into $w$-space as $\tilde{w}_i = q_{\tilde{z}_i}$. The integration path is then given by the straight line segments $1 \rightarrow \tilde{w}_1 \rightarrow \tilde{w}_2 \rightarrow \ldots \rightarrow w$.
As an optimisation, if $|w_i| > |\tilde{w}_i|$ (e.g., $w_2$ in figure~\ref{fig:path}), we can omit adding a node, as the straight line connecting the previous and following node cannot pass over the singular locus. In the case of the figure, the integration path is thus chosen to be $1 \rightarrow \tilde{w}_1 \rightarrow w$. Furthermore, nodes are only added for loci that are actually singular, as deformations across non-singular loci do not change the value of the integral. We note that care needs to be taken if the original integration path passes over a singular locus, i.e., $z_i = \tilde{z}_i$. In this case an endpoint of a line segment in $w$-space will coincide with a pole of the integrand. To evaluate such expressions, the integration path must be deformed to avoid the pole (see section~\ref{sec:usage} on how to specify deformations for our {\ginac} implementation). Based on this deformation, the node $\tilde{w}_i$ is shifted away from the pole.
The amount to shift is computed based on the surrounding nodes and poles, such that the contour runs far from any of the poles, which is useful for the numerical evaluation of the ordinary MPLs that will arise later.

At this point the eMPLs are written as iterated integrals of rational functions along integration paths consisting of straight line segments. These integrals can be evaluated in terms of rational functions and ordinary MPLs, integrated along the same paths. Using the same approach as in section~\ref{subsec:cutIntegrationPath}, these are rewritten as polylogarithms whose integration contours are straight lines starting at the origin. These objects can be evaluated to arbitrary precision using {\ginac}.

What remains is to estimate to which order each kernel needs to be expanded. We automatically perform the expansion to an appropriate order, such that we achieve the target accuracy set in {\ginac}. We implement the expansion order-by-order for each kernel, such that we can dynamically determine the required order.
The expansion proceeds in three steps. First, we expand each kernel to a starting order that is estimated based on an effective expansion parameter and the desired accuracy. This estimate is usually very close to the actual number of required orders. Next, we continue expanding each kernel individually, until the numerical effect of each order is smaller than the desired accuracy. Finally, we expand all kernels together until the numerical effect drops below the target accuracy. The last step is necessary in the case that each individual kernel has a small effect, but the sum of new orders of multiple kernels is larger.

We conclude this section with a simple explicit example. We consider the expansion of 
\begin{align}
    \Gt\left(\sm{2\\0};z,\tau\right) &= \int_0^z \diff t \, g^{(2)}(t,\tau),
\end{align}
with $\tau \in \cF$ and $z \in \cD'$.
The expansion of the integration kernel $g^{(2)}(z,\tau)$ reads
\begin{align}
    g^{(2)}(z,\tau) &= \left(2\pi i\right)^2 \left[ \frac{1}{12} - \sum\limits_{m=1}^\infty \left( q_z^{m} + q_z^{-m} \right) \frac{q_\tau^m}{\left(1-q_\tau^m\right)^2} \right],
\end{align}
where $q_z=\exp(2\pi i z)$ and $q_\tau=\exp(2\pi i \tau)$.
We then obtain with $w=\exp(2\pi i t)$
\begin{align}
    \Gt\left(\sm{2\\0};z,\tau\right)
    &= 2\pi i \int_1^{q_z} \frac{\diff w}{w} \left[ \frac{1}{12} - \sum\limits_{m=1}^\infty \left( w^{m} + w^{-m} \right) \frac{q_\tau^m}{\left(1-q_\tau^m\right)^2} \right] \nonumber \\
    &= 2\pi i \left[ \frac{1}{12} \ln(q_z) - \sum\limits_{m=1}^\infty \frac{1}{m} \left( q_z^{m} - q_z^{-m} \right) \frac{q_\tau^m}{\left(1-q_\tau^m\right)^2} \right].
\end{align}
For the numerical evaluation the infinite sum in the last expression is truncated at an appropriate order.
In this case, the coefficient of the $q_\tau^0$-term is a logarithm in $q_{z}$, while for $m \ge 1$ the coefficients of the $q_\tau^m$-terms are rational functions in $q_{z}$.

\section{Usage of Package}\label{sec:usage}

We have implemented our algorithm in {\ginac}~\cite{Bauer:2000cp}, a C++ library for computer algebra. Starting with version 1.8.10, the package is included in the standard {\ginac} installation, which can be obtained from \url{https://www.ginac.de}.

A general elliptic multiple polylogarithm
\begin{align*}
    \Gt\left(\sm{n_1&\dots&n_k\\z_1&\dots&z_k} ;z,\tau\right) \,,
\end{align*}
is represented in {\ginac} as
\begin{align}
    \text{\texttt{Gt(\{\{$n_1$,$z_1$\},\ldots,\{$n_k$,$z_k$\}\},$z$,$\tau$)}} \,. \label{eq:ginacsynatx}
\end{align}
The $n_i$ should be non-negative integers, while all other arguments may be arbitrary symbolic expressions.

If all arguments can be directly evaluated to numerical values, a \texttt{Gt} object may be evaluated using the \texttt{evalf} function. When compiled and executed, the following program
\begin{verbatim}
#include <iostream>
#include <ginac/ginac.h>
using namespace GiNaC;

int main() {
    ex expr = Gt(lst{lst{2,3-2*I/7},lst{1,-2+I/7}},1+I,2+3*I);
    std::cout << expr << " = " << expr.evalf() << std::endl;

    return 0;
}
\end{verbatim}
prints
\begin{align*}
    &\text{\texttt{Gt({{2,3-2/7*I},{1,-2+1/7*I}},1+I,2+3*I)}} \\
    &\qquad\text{\texttt{ = -0.24570758059458382605-7.8354094938915607015*I}} \,.
\end{align*}

Expressions containing multiple eMPLs can be evaluated using the \texttt{Gt::ex\textunderscore{}evaluate} function. This function is usually preferable over \texttt{evalf}, as it applies the algorithm to all eMPLs simultaneously. This enables reusing partial results and allows cancellations of intermediate terms.
This function also takes a list of replacement rules as an optional second argument. These replacements are applied to all \texttt{Gt} arguments prior to numeric evaluation, which allows keeping arguments symbolic for as long as possible. The syntax is illustrated in the following example:
\begin{verbatim}
symbol z{"z"}, tau{"tau"};
ex expr = Gt(lst{lst{1,numeric(1,6)+I/6}},z,tau)
        + 12*Pi*Gt(lst{lst{1,numeric(5,6)+I/6}},z,tau);
ex replacements = lst{z == numeric(1,3)+I/5, tau == numeric(2,5)+I/3};
std::cout << expr << " = "
          << Gt::ex_evaluate(expr, replacements) << std::endl;
\end{verbatim}

The function \texttt{Gt::lst\textunderscore{}evaluate} works similarly, except that its first argument is a list of expressions and it returns a list of numeric values. This allows reusing partial results across the multiple expressions.

For advanced usage, we further provide the functions \texttt{Gt::ex\textunderscore{}zisToFundamental}, \texttt{Gt::ex\textunderscore{}regularise}, \texttt{Gt::ex\textunderscore{}tauToFundamental} and \texttt{Gt::ex\textunderscore{}cutIntegrationPath} which perform steps~\ref{algo:step1} to \ref{algo:step4} of the algorithm separately, not including possible repetitions of earlier steps. We also provide \texttt{Gt::ex\textunderscore{}prepare}, which performs all preparation steps, and \texttt{Gt::ex\textunderscore{}qExpand}, which performs the $q$-expansion on fully prepared eMPLs. All of these functions take an expression as a first argument, optionally a replacement list as a second argument and return a transformed expression.

\paragraph{Usage in \texttt{ginsh}.}
The {\ginac} package comes with the interactive command prompt \texttt{ginsh}. The syntax for defining eMPLs in \texttt{ginsh} is identical to the one described above. The evaluation is performed using the function \texttt{Gt\textunderscore{}evaluate}, which corresponds to both the \texttt{Gt::ex\textunderscore{}evaluate} and \texttt{Gt::lst\textunderscore{}evaluate} functions from {\ginac}.\footnote{Since functions in \texttt{ginsh} do not have a definite return type, the distinction between \texttt{Gt::ex\textunderscore{}evaluate} and \texttt{Gt::lst\textunderscore{}evaluate} is not necessary.} This function always requires a (possibly empty) replacement list as a second argument. Interactive use of the package is illustrated in the following example:
\begin{verbatim}
> z = 1+I;
> tau = 2+3*I;
> expr = Gt({{2,z1},{1,z2}},z,tau);
> Gt_evaluate(expr, {z1 == 3-2*I/7, z2 ==-2+I/7});
-0.24570758059458382562-7.835409493891560702*I
\end{verbatim}
An overview of further provided \texttt{ginsh} functions is given in appendix~\ref{app:ginsh}.

\paragraph{Specifying deformations around poles.}
In section~\ref{sec:nqexpand} we mentioned that it may be necessary to specify deformations of the integration contour around poles in the integrand. Deformations can be specified as an optional third argument in the definition of each integration kernel $g^{(n)}(z,\tau)$ in expression in eq.~\eqref{eq:ginacsynatx}, i.e., \texttt{\{$n$,$z$,$\Delta$\}}, where $\Delta = \pm 1$. If the locus $z$ becomes singular and the integration contour passes over it, the path is deformed around $z$, as specified by $\Delta$.
A value of $\Delta = +1$ ($-1$) indicates that the path should pass the pole on the right (left) in the direction of integration, i.e., the path is deformed over the point $z \exp(-i \epsilon \Delta)$ for some appropriate value of $\epsilon > 0$.
For example, if the integration runs along the positive real axis, a value of $\Delta = +1$ leads to a deformation toward negative imaginary values. This convention is a natural extension of the corresponding convention for ordinary MPLs in {\ginac}~\cite{Vollinga:2004sn}.
If no deformation is specified, the default value $+1$ is used.

When deciding whether a pole lies on the integration path or merely close to it, it is important to avoid floating point errors.
Our implementation accepts input values in any form that can eventually be evaluated to numerical values. If inputs are passed in an exact form, e.g., as rational numbers or symbolic expressions, they are kept in this form for as long as possible. Since only rational transformations are applied to the inputs, the exact form can be retained throughout the entire preparation phase. Only during the final $q$-expansion are floating point values actually inserted. This avoids the accumulation of numerical errors throughout the algorithm. If inputs are passed as floating point values, this is no longer the case. Users are therefore advised to pass inputs in exact form whenever possible.
A pole is considered to lie on the integration contour, if its distance to the contour is less than the target precision of {\ginac}. If a pole is very close to the integration path, the user should ensure that the target precision is sufficiently high to resolve this distance, or to specify a deformation which moves the path to the correct side of the pole.

\paragraph{Performance considerations.}
The algorithm comes with certain global parameters that affect performance. The provided default values work well across the test cases described in section~\ref{sec:applications}. In certain cases, better performance may be achievable by adjusting these parameters. For long-running calculations, users are encouraged to experiment with different values.

The thresholds $r_\text{max}$ and $i_\text{max}$, introduced in section~\ref{subsec:cutIntegrationPath}, determine how close the integration path may get to the boundary of the fundamental domain $\cD'$ before being cut. They can be adjusted via the global variables \texttt{Gt::cutIntegrationPath\textunderscore{}threshold\textunderscore{}real} and \texttt{Gt::cutIntegrationPath\textunderscore{}threshold\textunderscore{}imag} respectively. The default value for both variables is $0.4$. For a given eMPL and target accuracy, adjusting these values might lead to a moderate performance improvement.
From within \texttt{ginsh}, these values can be adjusted using the function \texttt{Gt\textunderscore{}setCutThresholds($r_\text{max}$,$i_\text{max}$)}.

The global variable \texttt{Gt::enable\textunderscore{}tauToFundamental} can be set to enable or disable step 3 of the algorithm, the mapping of the modular parameter $\tau$ to its fundamental domain $\cF$. This step is not strictly necessary for obtaining the correct result, but it improves convergence behaviour. It is enabled by default and we find that in the majority of cases, it does improve performance significantly. In certain cases, it may be possible to achieve better performance by omitting this step.
This behaviour can be adjusted in \texttt{ginsh} using the function \texttt{Gt\textunderscore{}enableTauToFundamental($i$)}, where $i\in\{0,1\}$.

We investigate the runtime of different parts of our algorithm. We find that the preparation phase contributes negligibly and essentially the entire runtime is spent on the $q$-expansion. Within the $q$-expansion, the majority of the runtime is usually spent on the evaluation of MPLs. To speed up evaluations, we implemented a caching system that stores numerical values of MPLs and reuses them if they are encountered again. Even after implementing the cache, MPLs are responsible for a significant fraction of the runtime. For the Bhabha scattering example, discussed in the following section, typically more than 80\% of the total runtime is spent on the evaluation of MPLs.

\section{Applications}\label{sec:applications}

Currently there exists no public package that is capable of evaluating eMPLs for arbitrary complex arguments. Some packages are available that are capable of evaluating certain classes of eMPLs. In the following, we discuss existing options and their limitations and compare their performance against our algorithm.

Firstly, the \texttt{iterated\textunderscore{}integral} library, which is part of {\ginac} is capable of evaluating different iterated integrals, including eMPLs~\cite{Walden:2020odh}. This is done by rewriting the entire iterated integral as a series expansion around $z = 0$. This expansion has a limited region of convergence, which poses restrictions on the arguments of eMPLs that can be evaluated using this method. In particular, it requires $|z| < |z_i|$, $\im(z_j) < |\im(\tau)|$ and $\im(z-z_i) < |\im(\tau)|$ for every singular locus $z_i$, which poses an issue for practical applications.

Secondly, we investigate numerical integration by implementing the integration kernels $g^{(n)}(z,\tau)$ in \textsc{Mathematica} and using the \texttt{NIntegrate} function to evaluate the eMPLs. This approach does support arbitrary eMPL arguments and it allows the evaluation to high precision. Problems are expected for higher length eMPLs, as the numeric integration of iterated integrals with oscillating integrands fails to converge quickly.

We investigate two benchmark scenarios. For the first scenario, we pick four eMPLs of different lengths
\begin{align*}\begin{split}
    \mathmakebox[3cm]{\Gt\left(\sm{1      \\z_1            };z,\tau\right)} \\
    \mathmakebox[3cm]{\Gt\left(\sm{1&2    \\z_1&z_2        };z,\tau\right)} \\
    \mathmakebox[3cm]{\Gt\left(\sm{1&2&1  \\z_1&z_2&z_3    };z,\tau\right)} \\
    \mathmakebox[3cm]{\Gt\left(\sm{1&2&1&1\\z_1&z_2&z_3&z_4};z,\tau\right)}
\end{split}\hspace*{1cm}\text{with}\hspace*{-1.5cm}\begin{split}\begin{aligned}
    z_1 &= \tfrac{3}{8}+\tfrac{1}{3}i \,, & z &= \tfrac{1}{4}+\tfrac{1}{8}i \,, \\
    z_2 &= \tfrac{1}{3}-\tfrac{3}{7}i \,, & \tau &= i \,, \\
    z_3 &=-\tfrac{2}{5}+\tfrac{4}{9}i \,, \\
    z_4 &= \tfrac{6}{11} \,.
\end{aligned}\end{split}\end{align*}
These expressions are chosen such that they can be evaluated using the \texttt{iterated\textunderscore{}integral} library. We evaluate these expressions to different precision goals and measure the runtime.

\begin{figure}[!th]
    \centering
    \includegraphics[page=1]{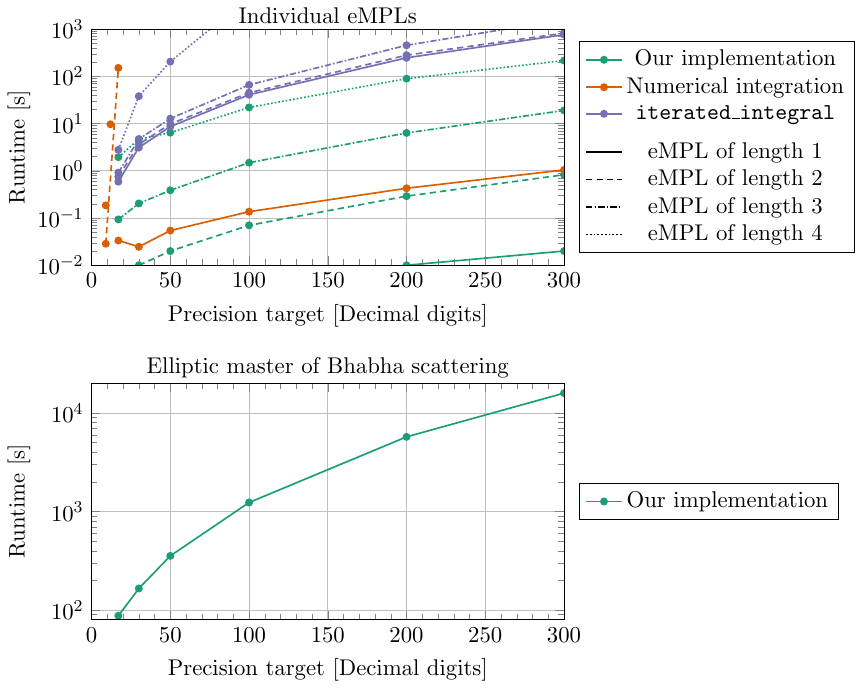}
    \caption{Runtime of the numerical evaluation of different eMPLs using our implementation, the \texttt{iterated\textunderscore{}integral} library and naive numerical integration using \texttt{NIntegrate} in {\sc Mathematica}, measured on a typical laptop. Upper plot: Four individual eMPLs of different lengths from 1 to 4. For lengths 3 and 4, only a single point could be computed using numerical integration. Lower plot: Elliptic master integral from two-loop Bhabha scattering.}
    \label{fig:runtime}
\end{figure}

As a second benchmark scenario, we have evaluated an elliptic planar master integral that appears in the two-loop QED corrections to Bhabha scattering where the full dependence on the electron mass is retained. In total, there are two planar and one non-planar integral family contributing to this process, all of which have by now been evaluated using differential equations~\cite{Henn:2013woa,Duhr:2021fhk,Delto:2023kqv}. It is known that the planar families can be expressed in terms of ordinary MPLs~\cite{Duhr:2021fhk,Heller:2019gkq}, and only the non-planar family requires the introduction of functions associated to an elliptic curve. However, the expression of some of the planar double-box integrals in terms of ordinary MPLs is extremely lengthy and cumbersome, rendering the numerical evaluation of these special functions inefficient. At the same time, it was shown that there is a compact expression of these integrals in terms of eMPLs~\cite{Duhr:2021fhk}. We have considered the double box integral from ref.~\cite{Duhr:2021fhk} (dubbed $\bar{f}$ in that reference), which depends on two dimensionless ratios $\tfrac{s}{m^2}$ and $\tfrac{t}{m^2}$, where $m$ is the electron mass. The analytic expression involves eMPLs of the form $\Gt\left(\sm{ 1   \\ z_1            }\cdots \sm{ 1   \\ z_l            };z,\tau\right)$, where $l\le 4$, and the singular loci are drawn from the set $\{\pm\tfrac{1}{8},\pm\tfrac{1}{4},\pm\tfrac{3}{8},0,\tfrac{1}{2},\tfrac{\tau}{2},\tfrac{1}{2}+\tfrac{\tau}{2}\}$. The kinematics is encoded in the point $z\in \mathbb{C}$ and the modular parameter $\tau$. In addition, the analytic expression also involves ordinary MPLs whose arguments are algebraic functions of $\tfrac{s}{m^2}$ and $\tfrac{t}{m^2}$. We have implemented this expression into \ginac, which allows us to evaluate all the special functions numerically.
We have evaluated this integral in the physical region where $s \sim M_Z^2$ and $t \lesssim 0$, which is the typical setup encountered for example when using Bhabha scattering to determine the luminosity of a lepton collider. Specially, our numerical input values of the special functions are
\begin{equation}
y = \tfrac{1}{2}(3-\sqrt{5}) = 0.382\ldots\,,\qquad z = 0.124\,,\qquad \tau = 0.877 i\,.
\end{equation}
We stress that, since we are working in the physical region, the amplitude develops an imaginary part. Some of these eMPLs lie outside the range of applicability of the \texttt{iterated\textunderscore{}integral} library, such that we cannot evaluate the expression using this library. Naive numerical integration is feasible for certain phase space points, but not for the specific point in the physical region that we chose. Therefore, we can evaluate this particular kinematic point only using our implementation in {\sc GiNaC}.
The results for the runtimes as a function of the requested number of digits are given in figure~\ref{fig:runtime}.

In summary, we find that the evaluation using naive numerical integration only scales well for eMPLs of length 1 and fails to converge quickly otherwise. The evaluation using \texttt{iterated\textunderscore{}integral} works well for arguments in the radius of convergence, also for eMPLs of higher lengths. It can, however, not be used to evaluate physical expressions such as the master integrals appearing in the two-loop QED corrections to Bhabha scattering. Our implementation allows for the efficient evaluation of all investigated eMPLs and precision targets. We conclude that our algorithm is well-suited for the evaluation of physically relevant eMPLs with arbitrary arguments and to high precisions and is the only available code to achieve this.

\section{Conclusions}\label{sec:conclusions}

In this work we have presented an algorithm to compute a convergent $q$-series of an eMPL. A distinctive feature of this $q$-series is the fact that its coefficients only involve ordinary MPLs. In this way, we can reduce the numerical evaluation of eMPLs to the evaluation of ordinary MPLs, which can be achieved efficiently and with high precision using existing public libraries.

We have also presented an implementation of our algorithm into the {\ginac} framework, which is already widely used by the particle physics community. Our package, which will be shipped together with {\ginac} starting from the version 1.8.10 allows for the efficient evaluation of eMPLs with arbitrary arguments, and it is the first public algorithm to achieve this. We find that the algorithm scales well to arbitrary precision and is therefore suited for both the numerical evaluation of physical observables involving eMPLs and the use of the PSLQ algorithm to find linear relations among eMPLs. We therefore foresee a vast range of possible future applications of our algorithm and its {\ginac} implementation, both in the context of Feynman integrals and to study eMPLs more generally from the mathematical side.

\section*{Acknowledgments}

We thank Thomas Gehrmann and Kay Schönwald for helpful discussions. RM and FL have been supported by the Swiss National Science Foundation (SNSF) under contract 240015 and by the European Research Council (ERC) under the European Union's Horizon 2020 research and innovation programme grant agreement 101019620 (ERC Advanced Grant TOPUP). RM was further supported by the European Union's Horizon 2020 research and innovation program EWMassHiggs (Marie Sk{\l}odowska Curie Grant agreement ID: 101027658) and the Emeritus Foundation. CD and SW are supported by the DFG Research Unit FOR 5582 ``Modern Foundations of Scattering Amplitudes''.

\appendix

\section{Convergence of the \texorpdfstring{$\bm{q}$}{q}-Expansion}\label{app:convergence}

In this appendix we determine convergence properties of the $q$-series in eq.~\eqref{eq:gQExpansion}. We consider the absolute convergence of the double sums appearing in eq.~\eqref{eq:gQExpansion} and ignore terms that do not appear in any sum. Then we only need to consider sums of the form

\begin{align}
    \sum_{m=1}^{\infty}  \text{trig}(2\pi m z) \sum_{n=1}^{\infty}n^{k-1} q_\tau^{mn} \,,
\end{align}
where $\text{trig}(x)$ acts as a representative of either $\sin(x)$ or $\cos (x)$.

A double sum $\sum_{i,j} a_{ij}$ converges absolutely if both $\sum_i b_i$ or $\sum_j c_j$ converge absolutely, where $b_i = \sum_j |a_{ij}|$ and $c_j = \sum_i |a_{ij}|$. We therefore consider the sum:
\begin{align}
    \sum_{m=1}^{\infty}  \left| \text{trig}(2\pi m z) \right| \sum_{n=1}^{\infty}n^{k-1} \left|q_\tau^{mn} \right|\,.
\end{align}
We perform the innermost sum over $n$. Since $\im(\tau) > 0$, we have $|q_{\tau}| < 1$, and the innermost sum converges. An explicit formula for the innermost sum is given in eq.~\eqref{eq:qSum}. Let us introduce the notation
\begin{align}
    b_m(k) \equiv \sum_{n=1}^{\infty}n^{k-1} \left|q_\tau^{mn} \right| \,.
\end{align}
We then need to analyse the convergence of the sum
\begin{align}
    \sum_{m=1}^{\infty}  \left| \text{trig}(2\pi m z) \right| b_m(k) \,.
\end{align}
We have
\begin{align}
    \left| \text{trig}(2\pi m z) \right| b_m(k)  \leq  \frac{1}{2}\left| \exp(2 \pi i z) \right|^m b_m(k) + \frac{1}{2} \left| \exp(2 \pi i z) \right|^m b_m(k) \,,
\end{align}
where we have expressed the trigonometric function $\text{trig}(x)$ through exponential functions and used the triangle inequality. Note that the convergence of an infinite sum only depends on its tail. We therefore consider the limit $m \rightarrow \infty$, where
\begin{align}
    b_m(k) \equiv |q_\tau^m| \frac{1+\cO(|q_\tau^m|)}{(1-|q_\tau^m|)^k} = |q_\tau^m| \left(1 + \cO\left(|q_\tau^m|^{2}\right) \right)  \rightarrow |q_\tau^m| \,.
\end{align}
In this limit, the tail of the resulting sums behaves as
\begin{align}
    \sum_{m=1}^\infty \left| \exp(\pm 2\pi i z)^m \right| |q_\tau^m|  = \sum_{m=1}^\infty \left| \exp(2 \pi i (\tau  \pm z)) \right|^m \,.
\end{align}
This sum converges if $\left| \exp(2 \pi i (\tau  \pm z)) \right| < 1$, which is exactly the case if $|\im(z)| < \im(\tau)$.

\section{Useful Functions in \texttt{ginsh}}\label{app:ginsh}
In the following table, we recap the functions available in \texttt{ginsh} for the user's convenience.
Functions marked with a star take two arguments: the expression to be evaluated and a list of replacements which assign numerical values to all variables appearing in the arguments of eMPLs.

\vspace*{\baselineskip}
\noindent
\begin{tabularx}{\textwidth}{l|X}
\texttt{Gt(\{\{$n_1$,$z_1$\},\dots,)\},$z$,$\tau$)}            & The eMPL $\tilde{\Gamma}\left(\sm{n_1&\dots&n_k\\z_1&\dots&z_k};z,\tau\right)$ with default treatment of poles on the integration contour. \\
\texttt{Gt(\{\{$n_1$,$z_1$,$\Delta_1$\},\dots)\},$z$,$\tau$)}  & The eMPL $\tilde{\Gamma}\left(\sm{n_1&\dots&n_k\\z_1&\dots&z_k\\\Delta_1&\dots&\Delta_k};z,\tau\right)$ where $\Delta_i = \pm1$ indicates in which direction to deform the integration contour around poles. \\
\texttt{Gt\_{}enableTauToFundamental($i$)}                     & Disable ($i=0$) or enable ($i=1$) step~\ref{algo:step3} of the algorithm as part of the \texttt{Gt\textunderscore{}evaluate} and \texttt{Gt\textunderscore{}prepare} functions. \\
\texttt{Gt\_{}setCutThresholds($r_\text{max}$,$i_\text{max}$)} & Set the thresholds which determine the domain $\cD'$ (cf.~eq.~\eqref{eq:fundamentalDomain_d2}). \\
\texttt{evalf($x$)}                                            & {\ginac}s default evaluation function. If all arguments are numeric, eMPLs in the argument $x$ will be evaluated sequentially using our algorithm. \\
\texttt{Gt\textunderscore{}evaluate}*                          & Evaluate an expression containing eMPLs. Intermediate results are reused between expressions. \\
\texttt{Gt\textunderscore{}prepare}*                           & Perform all preparation steps for the numeric evaluation of eMPLs and returns the prepared expression without performing the $q$-expansion. \\
\texttt{Gt\textunderscore{}zisToFundamental}*                  & Perform preparation step~\ref{algo:step1}. \\
\texttt{Gt\textunderscore{}regularise}*                        & Perform preparation step~\ref{algo:step2}. \\
\texttt{Gt\textunderscore{}tauToFundamental}*                  & Perform preparation step~\ref{algo:step3} (not including repetition of steps \ref{algo:step1} and \ref{algo:step2}). \\
\texttt{Gt\textunderscore{}cutIntegrationPath}*                & Perform preparation step~\ref{algo:step4} (not including repetition of step \ref{algo:step1}). \\
\texttt{Gt\textunderscore{}qExpand}*                           & Perform the $q$-expansion on fully prepared eMPLs.
\end{tabularx}

\bibliographystyle{JHEP}
\bibliography{biblio}

\end{document}